\documentclass[aps,prb,twocolumn,superscriptaddress,english,footinbib]{revtex4-1}

\usepackage[T1]{fontenc}
\usepackage{babel}
\usepackage{amsmath}
\usepackage{placeins}
\usepackage{amssymb}
\usepackage{wasysym}
\usepackage{graphicx}
\usepackage[caption=false]{subfig}
\usepackage{xcolor}
\usepackage{braket}

\usepackage[percent]{overpic}

\usepackage{multirow}

\usepackage[linktocpage=true,
  colorlinks=true, 
  pdfborder={0 0 0},
  linkcolor=blue,
  citecolor=red,
  filecolor=yellow,
  urlcolor=blue,
  bookmarks,
  pdfauthor={},
]{hyperref}

\newcommand{\sh}{SH$_3$}
\newcommand{\lah}{LaH$_{10}$}

\newcommand{\tc}{$T_\text{c}$}

\newcommand{\ep}{\textit{e-ph}~}

\begin{document}

\title{Phase Diagram and Superconductivity of Calcium Borohyrides at Extreme Pressures}
\author{Simone Di Cataldo} \email{simone.dicataldo@uniroma1.it}
\affiliation{Institute of Theoretical and Computational Physics, Graz University of Technology, NAWI Graz, 8010 Graz, Austria}
\affiliation{Dipartimento di Fisica, Sapienza Universit\`a di Roma, 00185 Roma, Italy} 
\author{Wolfgang von der Linden}
\affiliation{Institute of Theoretical and Computational Physics, Graz University of Technology, NAWI Graz, 8010 Graz, Austria}
\author{Lilia Boeri} \email{lilia.boeri@uniroma1.it}
\affiliation{Dipartimento di Fisica, Sapienza Universit\`a di Roma, 00185 Roma, Italy}

\date{\today}

\begin{abstract}
  Motivated by the recent discovery of near-room temperature superconductivity in high-pressure superhydrides, we investigate from first-principles the high-pressure superconducting phase diagram of the ternary Ca-B-H system, using ab-initio evolutionary crystal structure prediction, and Density Functional Perturbation Theory.
  We find that below 100 GPa all stable and weakly metastable phases are insulating.
  This pressure marks  the appearance of several new chemically-forbidden
  phases on the hull of stability, and the first onset of metalization in CaBH$_5$.
Metallization is then  gradually achieved at higher pressure at different compositions.
Among the metallic phases stable in the Megabar regime, we predict two high-\tc{} superconducting phases with CaBH$_6$ and Ca$_2$B$_2$H$_{13}$ compositions,
with critical temperatures of 119 and 89 K at 300 GPa, respectively, surviving to lower pressures.
Ternary hydrides will most likely play a major role in superconductivity research in the coming years;
our study suggests that, in order to reduce the pressure for the onset of metallicity and superconductivity, further explorations of ternary hydrides should focus on elements less electronegative than boron.
\end{abstract}

\maketitle

\section*{Introduction}

The search for high-temperature superconductors in hydrogen-rich systems stems from two seminal intuitions of Neil Ashcroft's, i.e. that hydrogen, under sufficiently high pressures, may be turned into a metallic room-temperature superconductor, and that the same result may be achieved at lower pressures exploiting \textit{chemical precompression} in hydrides \cite{Ashcroft_PRL_1968_metallic_H, Ashcroft_2004_PRL_H_dominant}.
This second intuition, in particular, led to a major breakthrough in 2014,
with the experimental discovery that sulfur hydride (\sh), under a pressure of 200 GPa,
becomes a superconductor with a critical temperature (\tc) of 203 K\cite{Drodzov_Nature_SH3, Duan_SciRep_2015_SH3}.
After the discovery of \sh, other superconducting hydrides were found, with lanthanum hydride (\lah) currently holding the record, with a \tc{} of 260 K \cite{Eremets_Nature_LaH10, Hemley_PRL_LaH10_exp, Oganov_arxiv_2019_YH6, Oganov_2020_mattod_ThH10, Eremets_arxiv_2015_PH3, Eremets_arxiv_2019_YH9}.
All these high-\tc{} hydrides are remarkable examples of high-pressure \textit{forbidden chemistry},
as their H-rich chemical compositions defy the standard rules on oxidation states and coordination;
it is arguable that their existence would have never been anticipated, without
theoretical predictions enabled
by \textit{ab-initio} methods based on Density Functional Theory (DFT) methods for crystal structure prediction and superconductivity,
developed in the last 20 years.\cite{Oganov_Xtal_drives, RevModPhys_DFPT_Baroni, AllenDynes_PRB_1975, Oliveira_PRL_1988_SCDFT, Sanna_JPSJ_2018_ME_theory,Boeri_PhysRep_2020}. 

\begin{figure}[h!t]
		\includegraphics[width = 0.45\textwidth]{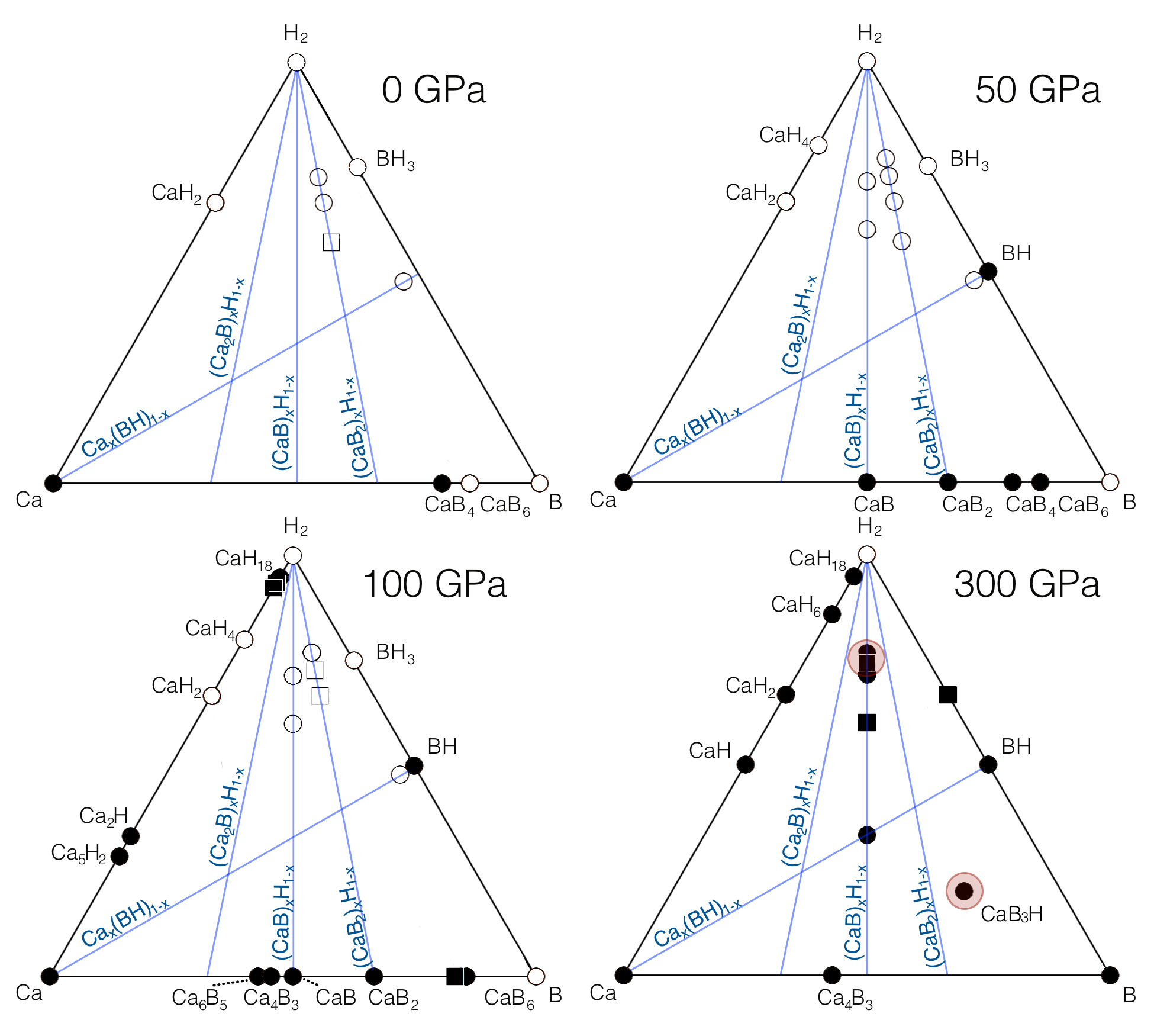}
	\caption{Convex hull diagrams for calcium, boron and hydrogen at 0 , 50, 100 and 300 GPa predicted from evolutionary crystal structure prediction calculations. Thermodynamically stable and metastable structures are shown as circles and squares, while empty and filled symbols represent insulating and metallic structures, respectively. Compositions within 50 meV/atom from the hull are considered metastable. Blue lines indicate lines where additional structural searches were carried out. The compositions circled in red - CaBH$_6$, Ca$_2$B$_2$H$_{13}$, and CaB$_3$H exhibit a finite \tc{}.}
	\label{fig:convexhulls}
\end{figure}

High-\tc{} superhydrides belong to the class of \textit{conventional} superconductors, where the superconducting pairing is mediated by phonons. For this class of materials, Migdal-Eliashberg theory, combined with first-principles calculations of the electron-phonon coupling, permits to make quantitative predictions of superconducting properties. Based on Migdal-Eliashberg theory and  experience,
the highest \tc{} among conventional superconductors are expected in materials, which:
$i$) contain light elements,  exhibit $ii)$ a sizable density of states (DOS) at the Fermi level, typical of good metals, and $iii)$ large electron-phonon (\ep{}) matrix elements,
characteristic of systems with contain directional (covalent or ionic) bonds \cite{SC:mgb2_Pickett_PRL2001,SC:diamond_Boeri_PRL2004}.
Whereas the first requirement is satisfied by any compound containing hydrogen,
which is the lightest element in the periodic table,
 the other two conditions are met only by  a few hydrides, and only at extreme pressures.

In the last five years,   many research groups, galvanized by the ground-breaking
\sh{} discovery,  devoted their efforts to the exploration of all possible binary hydrides, i.e. compounds with chemical formula A$_{x}$H$_{y}$.~\cite{Boeri_PhysRep_2020,oganov_review_hydrides,ma_review_hydrides,zurek_review_hydrides,pickard_review_hydrides}
As of 2020, this exploration has been essentially completed, showing that only around 10\% of the elements of the periodic table form high-pressure binary hydrides
with superconducting \tc{}s exceeding nitrogen boiling temperature (77 K). These can be grouped essentially in two main families, depending on the 
size and electronegativity of the \textit{A} element.

The first family is that of {\em covalent} hydrides, in which \textit{A} can be a metalloid or non-metal.
At low pressure, covalent hydrides form molecular crystals, which are large-gap insulators \cite{Parrinello_PRL_1996_H3O, Livas_EUPJB_2018_CH2, Stupian_JPCM_2007_CH4}. Pressures in the Megabar (100 GPa) range can render some of these compounds metallic, through band overlap. Some of these metallic molecular hydrides, such as phosphorous, silicon or sulfur
hydrides, exhibit superconductivity, with $T_{c}$s that can be as high as 100 K \cite{Eremets_Science_2008_SiH4, Ashcroft_PRL_2006_SiH4, Needs_PRL_2006_SiH4, Livas_2016_PRB_PH3, Eremets_arxiv_2015_PH3, Y_Ma_JCP_2014_H3S, Oganov_PRB_2017_GeHx, Ma_PRL_2008_GeH4, Cui_PCCM_2015_GeH4}.
Increasing pressure even further,  in a few systems molecular bonds break,
and, as  additional hydrogen is incorporated into the A-H lattice,
compounds with completely different stoichiometries and geometrical motifs form, 
in which hydrogen and the guest atom arrange in highly symmetric structures, which are metallic and held together by covalent, directional bonds. This is the case, for example, of SeH$_3$ and SH$_3$, where $T_{c}$'s can be as high as 200 K \cite{Livas_EPJB_2016_H3Se, Drodzov_Nature_SH3, Duan_SciRep_2015_SH3, Errea_PRL_SH3_2015, Heil_PRB_2015_H3S, Arita_PRL_2016_SH3}.

The second family of high-\tc{} binary hydrides comprises \textit{weak} hydrogen formers, such as alkaline earths, lanthanoids, actinoids, and early transition metals. Within this family, the highest \tc{}s are usually found in hydrogen clathrate structures, characterized by highly symmetric H cages surrounding a central atom \cite{MA_PRL_2017_clathrates}. Several of these clathrate hydrides, which form at pressures around 100 GPa, were predicted to achieve superconducting $T_{c}s$ above 200 K, \cite{Ma_PNAS_2011_CaH6, Wang_RSC_2015_MgH6,  MA_PRL_2017_clathrates, Heil_PRB_2019_YH6, Ashcroft_2017_PNAS_clathrates, Oganov_2020_mattod_ThH10}, and YH$_6$, YH$_9$, LaH$_{10}$, ThH$_{10}$ were experimentally confirmed as near-room temperature superconductors \cite{Hemley_PRL_LaH10_exp, Eremets_Nature_LaH10, Eremets_arxiv_2019_YH9, Oganov_arxiv_2019_YH6, Oganov_2020_mattod_ThH10}.

After the binary hydrides have been thoroughly explored, the natural route to improve the superconducting properties of high-pressure hydrides, either by reducing the pressures needed to achieve high \tc{}, or by increasing the maximum \tc{}, is to consider ternary hydrides - materials with A$_{x}$B$_{y}$H$_{z}$ composition, where the presence of two different elements, \textit{A} and \textit{B}, opens many routes for the optimization of material properties.
However, the increased flexibility of ternary hydrides comes at a high price, since accounting for possible decomposition paths requires exploring a much larger compositional space than in the case of binaries.
For this reason, only a few out of the  six-thousand possible ternary hydrides
have been investigated to date \cite{ptable_footnote}. 
Most studies in literature have been focusing on optimization of known hydrides, by iso- or eterovalent substitution, \cite{Cui_FrontPhys_2018_H3SXe, Hemley_PRB_2020_CH4_H3S, Yao_PRB_2016_H3S_P, Li_PRB_2018_H_S_Se, Heil_PRB_2015_H3S, Amsler_PRB_2019_S_Se_H, Liang_PRB_2019_LaSH, Bergara_PRB_2019_CaYH12}, and only a few works taken into
 account the full thermodynamics of ternary phases\cite{Kokail_PRB_2017,MA_PRL_2019_LiMgH}. 

In this work, we used first-principles methods for evolutionary structure prediction~\cite{CompPhysComm_Oganov_2006_USPEX} and superconductivity \cite{RevModPhys_DFPT_Baroni, JPCM_Giannozzi_2009_QE} to explore computationally the full high-pressure phase diagram of the calcium-boron-hydrogen (Ca-B-H) ternary system to identify new superconductors.
Calcium boron hydrides are part of the broad family of metal-boron-hydrides (MBH), which have been extensively studied for hydrogen storage applications at room pressure.~\cite{Zuettel_ScrMat_2007_BH4, Jensen_ChemSocRev_2017_M_Borohydr}. MBH are characterized by the simultaneous presence of a weak (M) hydride former, and a strong (B) hydride former, which forms strong (covalent or ionic) bonds with hydrogen. At ambient conditions, metal borohydrides adopt a M(BH$_4$)$_n$ stoichiometry,
in which tetrahedral BH$^{-}_{4}$  anions, similar to methane, are loosely arranged on an open lattice, and accomodate  mono-, di- or trivalent cations in the interstitials.  Besides the stable M(BH$_4$)$_n$  compositions, other metastable phases have been predicted to form,\cite{Wolverton_PRB_CaB2H6} containing BH$_3^{-}$ and BH$_2^{-}$ anions, with ethane- and polyethylene-like motifs. 
The similarity between MBH and hydrocarbons makes them very attractive candidates for superconductivity, since they combine in a single compound the strong directional bonds typical of 
covalent hydrides,\cite{Livas_EUPJB_2018_CH2} with the possibility of controlling charge doping via interstitial atoms. Indeed, a theoretical study in lithium boron hydride showed that, while at ambient pressure all stable and metastable phases are wide-gap ionic insulators, at pressures in the megabar
range  covalent metallic structures with increased hydrogen content should become stable; a high-\tc{},
metallic, highly symmetric Li$_2$BH$_6$ phase with a \tc{} of 98 K was predicted to form at 100 GPa\cite{Kokail_PRB_2017}.

In this paper, we will analyze the high-pressure phase diagram of Ca-B-H with the precise
aim of identifying new hydrogen-rich, covalent metallic structures, analogous to \sh, 
exhibiting high-\tc{} conventional superconductivity. We will show that, although such structures do
indeed form,  electronic structure features amenable to the chemical properties of calcium and boron 
limit the maximum \tc{} achievable and their stability range, making calcium borohydrides not competitive with the record binary hydrides, such as \sh{} and \lah.

The paper is organized as follows: in section \ref{sect:phase_diagram} we will present the phase diagram as a function of pressure, highlighting the evolution of the structural properties of the stable phases. In section \ref{sect:elec_struct} we will present the electronic properties of the stable and metastable phases, focusing in more detail on the new metallic structures that we identify at high pressures. In section \ref{sect:supercon}, we will then investigate the electron-phonon properties, and discuss which specific electronic and structural features lead to low or high-\tc{} in the different phases.
In section \ref{sect:conclusions} we will draw the main conclusions of our work.
The Appendixes contain figures of the electronic DOS for all structures,
band structure plots for selected high-pressure phases, as well as an extensive account of the computational details.

\begin{figure*}[ht]
	\centering
	\includegraphics[width = 1\textwidth]{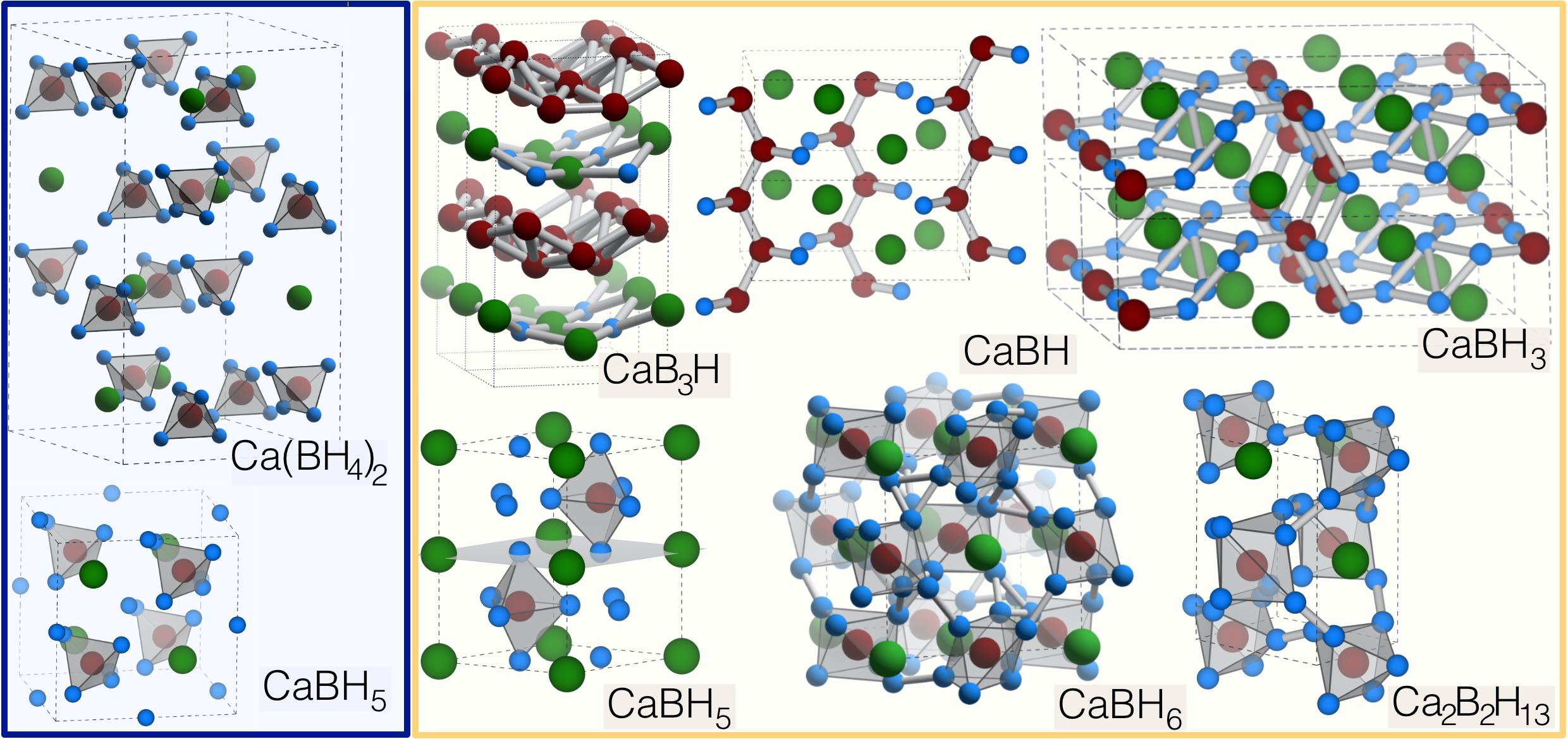}
	\caption{
          Left: crystal structures of the $Fddd$ phase of Ca(BH$_4$)$_2$ at 0 GPa,  and $F\bar{4}3m$ of CaBH$_5$ at 50 GPa. Right: crystal structures of high-pressure $Amm2$ phase of CaB$_3$H, $Ima2$ phase of CaBH, $C2/m$ phase of CaBH$_3$, $P63mmc$ phase of CaBH$_5$,  metastable $Pa\bar{3}$ phase of CaBH$_6$, and $Pm$ phase of Ca$_2$B$_2$H$_{13}$. Calcium, boron, and hydrogen are shown as green, red, and turquoise spheres, respectively.}
	\label{fig:structures_collected}
\end{figure*} 

\section{Phase diagram}
\label{sect:phase_diagram}

The Ca-B-H phase diagram at ambient pressure is well-characterized, since it was extensively studied for hydrogen storage applications. The ground state phase Ca(BH$_4$)$_2$ decomposes above 300$^o$C through different channels: either towards CaH$_2$, H$_2$, and CaB$_{12}$H$_{12}$, or towards CaH$_2$, CaB$_6$ and H$_2$,  through an intermediate Ca(BH$_3$)$_2$ phase\cite{Hauback_JMC_2008_CaB2Hx, Jena_ChemComm_2015_CaB2H6, Cho_JPCC_2012_CaB2Hx, Cho_JPCC_2012_CaB2Hx_backpressure, Wolverton_JACS_2008_CaB12H12}. These experimental observations are reproduced by first-principles calculations, which showed that the two pathways are quasi-degenerate \cite{Wolverton_PRB_CaB2H6}.

To the best of our knowledge, there are no predictions on the behavior of the ternary phase diagram under pressure, including intermediate Ca-B-H compositions. However, the pressure behavior of the binary systems (Ca-B, Ca-H, B-H), which form the edge of the ternary hull, was studied by \textit{ab-initio} calculations in recent years, in search for possible superconductors. Superconducting structures were found in all three systems, with a maximum \tc{} of 235 K in CaH$_6$ and 21 K in BH, both above 150 GPa, and 6 K in CaB at 30 GPa \cite{Kolmogorov_PRB_2013_CaB, Ma_PNAS_2011_CaH6, Oganov_PRL_2013_BH}.

In this study, to obtain the phase diagram as a function of pressure, 
we sampled the full ternary Ca-B-H phase space 
with variable-composition structural searches at fixed pressures (0, 50, 100, and 300 GPa) to identify the compositions stable at room, intermediate, and extreme pressures. More than 4000 unique structures were sampled for each pressure. The lowest-enthalpy structures for each composition identified in these runs were then relaxed at intermediate pressures to obtain the phase diagram. Further details can be found in Appendix \ref{app:ComputationalDetails}.

Fig.\ref{fig:convexhulls} shows the ternary convex hulls for the four pressures studied. The stable and metastable compositions are shown as circles and squares, respectively, while metal/nonmetal character is indicated by full/empty symbols. The figures highlight a gradual transition 
from low to high-pressure stoichiometries, and from insulating to metal behavior. 

At ambient pressure, our predictions agree with previous experiments and calculations \cite{Wolverton_PRB_CaB2H6, Kolmogorov_PRB_2013_CaB, Ma_PNAS_2011_CaH6, Oganov_PRL_2013_BH, Wolverton_JACS_2008_CaB12H12, David_PRB_2011_CaB2H8, Cui_JPCC_2018_CaB2H8}.  At room pressure the stable (or weakly metastable) phases lie along the Ca(BH$_x$)$_2$ line. Their structures comprise BH$^{-1}_{x}$ anions, with hydrocarbon-like motifs, and Ca cations in the interstitials: Ca(BH$_2$)$_2$ (polyethylene), Ca(BH$_3$)$_2$ (ethane), and Ca(BH$_4$)$_2$ (methane). In addition, there is a stable CaB$_{12}$H$_{12}$ phase, containing B$_{12}$H$_{12}$$^{2-}$ icosahedra \cite{Udovic_JSSC_2010_CaB12H12, Wolverton_JACS_2008_CaB12H12}. 
The crystal structure of Ca(BH$_4$)$_2$ is shown in the top left panel of Fig.\ref{fig:structures_collected}. 

 At 50 GPa, most compositions stable at ambient pressure remain on the convex hull, with 
 structures which maintain the motifs observed at ambient pressures. In addition,  new compositions appear along the Ca(BH$_x$)$_2$ and
 Ca(BH$_x$) lines, as hydrogen is trapped in the open molecular structures stable at room pressure.  
In Ca(BH$_5$)$_2$   interstitial H$_2$ molecules are trapped between BH$_4$ tetrahedra, in $P2_1/m$ CaBH$_3$ a single H atom occupies the interstitial sites among Ca tetrahedra, which are sandwiched between BH$_2$ linear chains, 
while in  $F\bar{4}3m$ CaBH$_5$  a single H atom is trapped in the interstitial sites
between  BH$_4$ anionic tetrahedra -- bottom left panel of Fig.\ref{fig:structures_collected}.  
Trapping of atomic or molecular hydrogen is observed at high pressures also in many covalent hydrides.~\cite{Duan_SciRep_2015_SH3}
 
 At 100 GPa, phases on the Ca(BH$_x$)$_2$ line become unstable, while CaBH$_3$ and CaBH$_5$, on the (CaB)H$_x$ line, remain stable. In addition, CaB$_{12}$H$_{12}$ has a structural phase transition towards a phase with $R\bar{3}m$ space group, in which the B$_{12}$H$_{12}$$^{2-}$ icosahedra are more densely packed. All structures are still insulating at this pressure, but metallization sets in by band overlap immediately after in CaBH$_5$
 
None of the Ca(BH$_x$)$_2$ ambient-pressure compositions survives up to 300 
GPa, where the pressure is so high that it leads to a complete rearrangement of the bonds:
 boron and  hydrogen now form interconnected, dense networks, and the structures are all metallic.
The stable high-pressure crystal structures are shown in the right panel of Fig. \ref{fig:structures_collected}.

 CaB$_3$H is characterized by buckled triangular boron layers, similar to those predicted in Ref.\onlinecite{Oganov_PRL_2013_BH} for pure boron, alternated with CaH planes, so that the Ca atoms fits into the valleys of the B layers.  CaBH is constituted by infinite B-H buckled chains in the $z$ direction, parallel to Ca chains.  
 Ca$_2$B$_2$H$_{13}$ exhibits a structure with distorted B-H octahedra and triangular bipyramids, and interstitial H$_2$ molecules.
 CaBH$_5$ undergoes a structural phase transition from the $F\bar{4}m$ phase identified at 50 GPa and shown in the left panel of Fig.~\ref{fig:structures_collected},
 to a phase with space group $P6_{3}mmc$, in which the interstitial H atoms are incorporated in the B-H sublattice to form 
  BH$_5$ triangular bipyramids.

  In addition to these thermodynamically stable phases, among H-rich compositions we also identified a  high-symmetry structure with space group $Pa\bar{3}$ for CaBH$_6$, lying about 100 meV/atom above the hull. This structure comprises BH$_6$ 6-vertex antiprisms, with hydrogen at the vertexes and boron in the middle. The  nearest-neighbor H-H distance (1.2\AA) is close to that of atomic hydrogen,~\cite{Ceperley_PRL_1993_atomic_H}, suggesting the formation of a metallic H sublattice.
    Given the empirical correlation between H-H interatomic distances and high-\tc{} superconductivity
  in high-pressure hydrides, we decided to include this structure in our pool of potential high-\tc{}
  superconductors.
 
 \begin{figure}[ht]
 	\includegraphics[width = 0.48\textwidth]{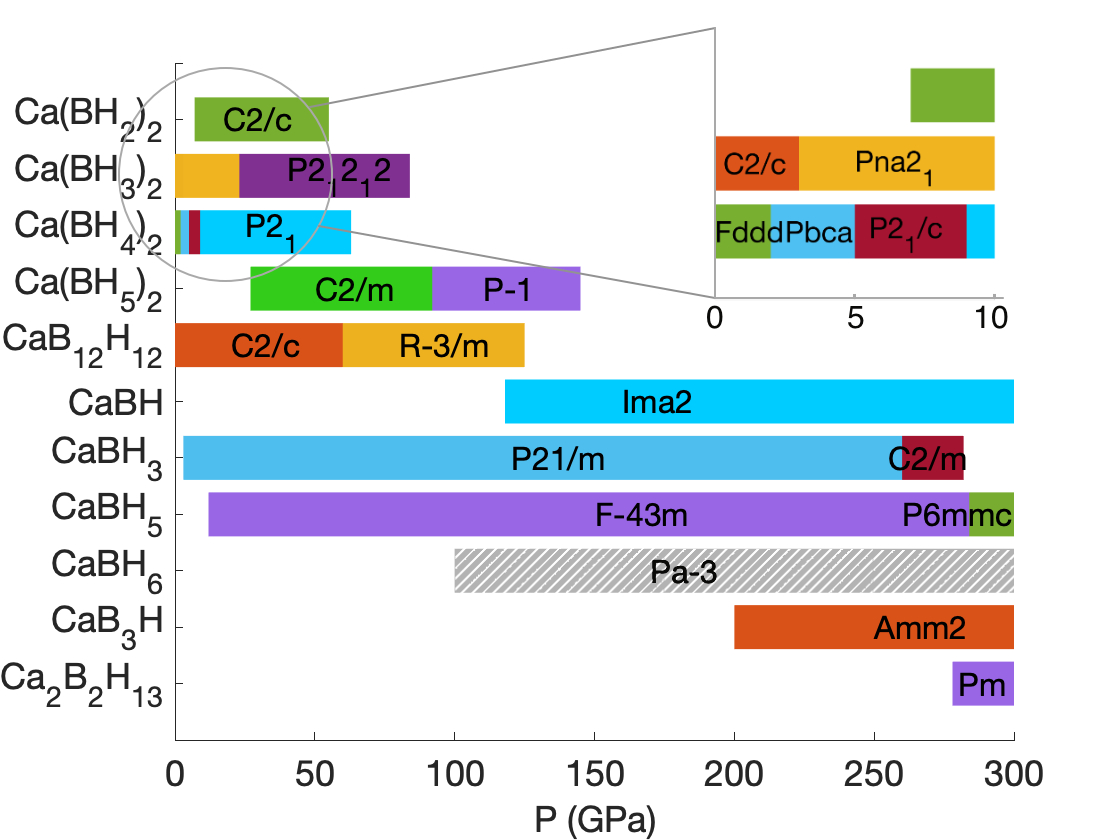}
 	\caption{Phase diagram for intermediate compositions as a function of pressure. The space group for each phase is indicated on top of the bars.
          The slashed pattern indicates the dynamical stability of a high-symmetry CaBH$_6$ structure, which is thermodynamically unstable. The inset shows phase transitions between low-pressure structures
        for Ca(BH$_x$)$_2$ compositions.}
 	\label{fig:phase_diagrams_bars}
 \end{figure}

The phase diagram in 
Fig.~\ref{fig:phase_diagrams_bars} summarizes the stability range of the different
Ca-B-H structures as a function of pressure. As it can be appreciated, several phases which appear
on the 300 GPa convex hull remain stable up to much lower pressures.

There is a rather clear separation of the phase diagram between a low- and a high-pressure regime, setting in at around 100 GPa.
The group of low-pressure phases comprises Ca(BH$_x$)$_2$ (x = 2, 3, 4) compositions, 
as well as CaB$_{12}$H$_{12}$; as shown previously, all these phases form molecular, insulating crystals. A series of structural transitions occurs between 0 and 50 GPa, due to the progressive rearrangement of BH$_x$ units in the crystal.
At pressures beween 10 and 20 GPa, hydrogen starts to get trapped in
molecular or atomic form in these open structures, stabilizing H-rich stoichiometries, such as
Ca(BH$_5$)$_2$, Ca(BH$_3$) and Ca(BH$_5$). 

Additional structures along the Ca(BH)$_x$ line become stable above 100 GPa: in these phases, the 
low-P motifs of the B-H sublattice, analogous to hydrocarbons, are replaced by new ones, in which boron is coordinated to five or more hydrogen atoms. This qualitative transition of the B-H sublattice
behavior is accompanied by the onset of metallic behavior. In fact, all structures we report at 300 GPa are metallic in their whole stability range.

\section{Electronic Structure}
\label{sect:elec_struct}
\begin{figure}[htpb] 
	\centering
	\includegraphics[width = 0.45\textwidth]{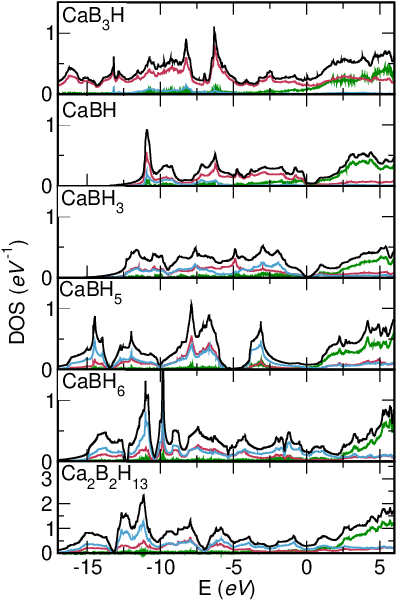}
	\caption{Calculated total and projected DOS for stable and metastable calcium boron hydrides at 300 GPa. Black, red, turquoise, and green lines represent the total DOS, and its projection onto B, H, and Ca, respectively. The energy range is shifted with respect to the Fermi level. The DOS is in units of states per eV per formula unit.
From top to bottom:
          $Amm2$ - CaB$_3$H, $Ima2$ - CaBH, $C2/m$ - CaBH$_3$, $P6_3mmc$ - CaBH$_5$, $Pa\bar{3}$ - CaBH$_6$ , and the $Pm$ - Ca$_2$B$_2$H$_{13}$.
	}
	\label{fig:eledos}
\end{figure}
We now discuss the electronic properties of the structures in the Ca-B-H high-pressure phase diagram. Structures which are stable up to 100 GPa are of no interest for superconductivity, since they are insulating.
Hence, here we discuss the main features of their electronic structure only briefly, and
collect the plots of the relative Densities of States (DOS) in Appendix \ref{app:supp_figures}.

Room pressure structures, characterized by a disconnected lattice of hydrocarbon-like, BH$^{-}_{x}$ anions, exhibit molecular-like DOS's, with sharp peaks of mixed B and H character, and band gaps ranging from 2.6 eV for Ca(BH$_2$)$_2$ to 5 eV for Ca(BH$_4$)$_2$. Ca $4s$ and $3d$ states lie at the bottom the conduction band; Ca is fully ionized, as confirmed by a Bader charge analysis, which assigns a total charge of about +1.5 $e$ to it. As pressure is increased  to 50 GPa, the peaks of the DOS broaden and  merge into a continuous band;  the bonding/antibonding gaps  decrease to 1.8 eV for Ca(BH$_2$)$_2$, and 3.4 eV for Ca(BH$_4$)$_2$.  The new Ca(BH$_5$)$_2$, CaBH$_3$ and CaBH$_5$ phases are all insulating, with gaps of 4.1, 1.6, and 2.3 eV, respectively.

At 100 GPa the reported phases for Ca(BH$_3$)$_2$, Ca(BH$_4$)$_2$, Ca(BH$_5$)$_2$ and CaB$_{12}$H$_{12}$ are still insulating, with calculated gaps of 1.6, 0.5, 1.8 and 2.6 eV, respectively, CaBH$_3$ exhibits a gap of 0.4 eV,
CaBH$_5$ is semimetallic.  

As shown in Fig.~\ref{fig:phase_diagrams_bars}, phases resulting from a the full rearrangement of the B-H sublattice at high pressures 
gradually appear on the convex hull only above 100 GPa. We 
discuss their electronic properties at a common pressure of 300 GPa, where high-P structures are stable for all compositions.
 In Fig.\ref{fig:eledos} we show the electronic DOS projected onto the atomic orbitals 
 for CaB$_3$H,  CaBH, CaBH$_3$, CaBH$_5$, CaBH$_6$, and Ca$_2$B$_2$H$_{13}$. The
 energies have been rescaled with respect to the Fermi energy.
The behavior of the DOS in the valence region depends strongly on the relative boron and hydrogen content: the CaB$_3$H and CaBH DOS are dominated by electronic states with B character. 
On the other hand, the hydrogen-rich phases exhibit a strong B and H hybridization over the whole -20 to 5 eV range;  furthermore, Ca-$4s$ states are spread over the valence and conduction band, so that  Ca is only partly ionized, with an average net Bader charge of +0.9.
In CaBH$_5$ the valence band exhibits B and H character with the exception of energies near the Fermi level, where around half of the contribution to the total DOS comes from states with Ca-$3d$ character. 

In CaBH$_6$ and Ca$_2$B$_2$H$_{13}$, the occupied states mostly have B and H character; the B- and H- projected DOS follow each other rather closely, which indicates that the two are covalently bonded, while the Ca-$3d$ states remain empty, and lie around 2 eV above the Fermi energy.

\section{Superconductivity}
\label{sect:supercon}
\begin{figure}[htpb]
	\centering
	\includegraphics[width=0.95\linewidth]{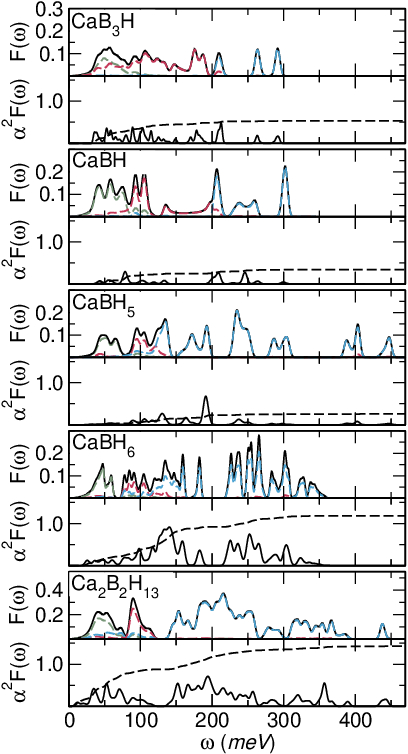}
	\caption{\label{fig:a2f_phonons} Atom-projected phonon DOS ($F(\omega)$), Eliashberg function ($\alpha^2F(\omega)$, solid lines), and frequency-dependent \ep{} coupling constant ($\lambda(\omega)$, dashed lines)
calculated at 300 GPa for the same six high-pressure phases as in Fig.~\ref{fig:eledos}.
The total phonon DOS is shown in black, and its projections on Ca, B, and H, are shown in green, red, and blue, respectively. }
\end{figure}

With the exception of CaBH$_3$,  which is a semimetal with a very small DOS at the Fermi level,
all high-pressure Ca-B-H phases are metallic, and hence  have the potential to be high-\tc{} superconductors.
In order to quantitatively assess their superconducting properties, we carried out calculations
of their electron-phonon properties, using Density Functional Perturbation theory, as implemented in the
plane-waves pseudopotential code \textit{Quantum Espresso}.~\cite{Savrasov_PRB_1996_LRT, JPCM_Giannozzi_2009_QE, RevModPhys_DFPT_Baroni}. 

The top panels of  Fig.\ref{fig:a2f_phonons} show the calculated total and partial phonon Densities of States of CaB$_3$H, CaBH, CaBH$_5$, CaBH$_6$, and Ca$_2$B$_2$H$_{13}$, with the same
color coding as for the electronic DOS's in Fig.~\ref{fig:eledos}.

For all structures there is a net separation of phonon spectra in three regions:
a low-energy one, extending from 0 to 75 meV, dominated by modes of Ca character, an intermediate one, from 75 to 175 meV,
with modes of mixed B and H character, and a high-energy region, from 175 up to 300 meV, characterized by B-H stretching modes.
Moreover, hydrogen-rich structures exhibit a further high-energy band, above
300 meV, which involves purely H vibrations and is not present in CaB$_3$H and CaBH. 
%

The bottom panels of the figure show, for each Ca-B-H phase, the calculated Eliashberg (electron-phonon) spectral function, which
is essentially a phonon Density of States, weighted by the electron-phonon matrix
elements for states at the Fermi level.~\cite{Carbotte_RevModPhys_1990}
Comparing its shape with the phonon DOS one can obtain a direct information on the distribution of the \ep coupling over the phonon spectrum.
In all cases, the Eliashberg function follows quite closely the phonon DOS,  indicating that the \ep coupling is distributed almost evenly on all vibrational modes, in analogy with other high-\tc{} high-pressure hydrides \cite{Duan_SciRep_2015_SH3, MA_PRL_2017_clathrates, Errea_Nature_2020_LaH10, Heil_PRB_2019_YH6}.

From the Eliashberg function, we obtained the logarithmic average phonon frequency
$\omega_{log}$ and the \ep{} coupling constant $\lambda$, which appear in the Mc-Millan
Allen Dynes formula for the critical temperature:\cite{AllenDynes_PRB_1975, McMillan_PR_1968}
\begin{equation}
T_c = \frac{\omega_{log}}{1.2}\exp\left( \frac{-1.04(1+\lambda)}{\lambda(1-0.62\mu^{*})-\mu^{*}} \right)
\label{eq:allendynes}
\end{equation}
and measure respectively the average effective frequency of the phonons participating in
the superconducting pairing and the intensity with which they couple to electrons at the Fermi level.

These  quantities are reported in table \ref{tab:tcdostable} for all five Ca-B-H compounds shown in Fig.\ref{fig:a2f_phonons}. The last column of the table shows also the \tc{}s, obtained
 from Eq.~\ref{eq:allendynes} for a standard choice of the Morel-Anderson pseudopotential, $\mu^*=0.1$.
\begin{table*}[h!tbp]
	\begin{tabular}{ccccccccccc}
		\hline
	& \begin{tabular}{@{}c@{}}P \\ (GPa) \end{tabular}
	& \begin{tabular}{@{}c@{}}  space group \\  \end{tabular}  
	& \begin{tabular}{@{}c@{}}  f.u. \\  \end{tabular}  
	& \begin{tabular}{@{}c@{}} Volume \\ (\AA$^{3}$) \end{tabular} &
	\begin{tabular}{@{}c@{}}$N(E_F)$\\ (10$^{2}$ $\cdot$ st/ eV \AA$^{3}$)\end{tabular} &	
	$N_H/N(E_F)$ & $\lambda$ & 
	\begin{tabular}{@{}c@{}}$\lambda/N(E_F)$  \\ (10$^{-2}$  eV \AA$^{3}$)\end{tabular}  &     
	\begin{tabular}{@{}c@{}}$\omega_{log}$ \\ (meV) \end{tabular} 	
	& \begin{tabular}{@{}c@{}} \tc{} \\ (K) \end{tabular} 	\\

		\hline
		\hline
		CaB$_3$H        &	300	&	$Amm2$	& 2	& 43.6		&	1.08 &	0.02	& 0.48 &0.44	& 78	&	7 	 	\\
		CaBH		  &	300		&	$Ima2$	& 2	& 27.9	&	0.60 &	0.06	& 0.14&0.23 	& 97	&	$\leq$ 0.1		\\
		CaBH$_5$      &	300	&	$P6mmc$	& 2	& 39.9		&	0.41&	 0.40	& 0.26&0.63 	& 131	&	$\leq$ 0.1		\\
		CaBH$_6$      &	100	&	$Pa\bar{3}$	& 4 	&	119.7		&	0.57	& 	0.73			&	 1.93			&3.33		&70			&114				\\		
		CaBH$_6$      &	200	&	$Pa\bar{3}$	& 4	& 	97.6		&	0.64	&	0.67			&		1.27		&1.99		&106		&117				\\
		CaBH$_6$      &	300	& $Pa\bar{3}$		& 4	&  85.6			&	0.79		&	 0.67	& 1.19	&1.59      & 117	&	119		\\
		Ca$_2$B$_2$H$_{13}$	& 200	& $Pm$  & 1	& 50.2&	0.59 &	 0.51		& 1.23&2.10	     & 59	&	63		\\
		Ca$_2$B$_2$H$_{13}$	&300	&   $Pm$	& 1	& 43.9 &	0.67 &	 0.55	& 1.37&2.07	     & 74	&	89		\\

		\hline
	\end{tabular}
	\caption{Normal and superconducting state properties of the six high-pressure Ca-B-H phases in Fig.~\ref{fig:eledos}-\ref{fig:a2f_phonons}.
          Space group, number of formula units, unit cell volume; 
          DOS at the Fermi level - $N(E_F)$, Hydrogen fraction of the total DOS at the Fermi level ($N_H/N(E_F)$), electron-phonon coupling constant ($\lambda$), effective \ep{} matrix element ($\lambda/N(E_F)$),
          logarithmic-average phonon frequency ($\omega_{log}$), McMillan-Allen-Dynes superconducting \tc{} -- Eq.~\ref{eq:allendynes}, with $\mu^*=0.1$.
          The DOS at the Fermi level is expressed in states/spin, and is rescaled by volume to allow for an easier comparison between different pressures
	}
	\label{tab:tcdostable}
\end{table*}

For only two out of five structures we predict a \tc{} exceeding liquid nitrogen boiling point: CaBH$_6$ (\tc=119~K), and Ca$_2$B$_2$H$_{13}$ (\tc=89~K). Both
are hydrogen-rich, high-symmetry phases.
In both cases, superconductivity survives up to lower pressures, with sizable \tc{}s.

 Boron-rich CaB$_3$H also exhibits a finite, although
much lower \tc (7 K). This is not surprising since an inspection of the electronic DOS
in fig: \ref{fig:eledos} shows that the contribution of H states is negligible in the whole valence region,
and the fraction of H states at the Fermi level, $N_H/N_{tot}$, is as low as 2\%. Essentially, the superconducting pairing is  dominated by electronic and vibrational states of the boron sublattice, determining an intermediate \ep{} coupling constant ($\lambda=0.48$), and an $\omega_{log}$  which is sensibly larger than in superconducting borides, but about half than high-\tc{} hydrides.~\cite{Boeri_PhysRep_2020}

The other two metallic  calcium borohydrides stable at high-pressure, CaBH and CaBH$_5$,
should not be superconducting according to our calculations.
Both exhibit a DOS at the Fermi level which is comparable to that of other high-pressure hydrides;
note that the value of the DOS in table \ref{tab:tcdostable} has been rescaled with the unit cell volume,
in order to allow for a meaningful comparison across structures with different compositions and stability pressures.

However, the value of the parameter $\lambda/N(E_F)$,
 is remarkably smaller than in the two high-\tc{} Ca-B-H phases. This parameter essentially
 measures the intrinsic coupling between electrons at the Fermi level and lattice vibrations.
In CaBH, its low value can be explained in terms of a very small fraction of H-derived electronic
states at the Fermi level ($N_H/N_{tot}=0.06$), leading to ineffective \ep{} coupling.
The same argument cannot be applied in CaBH$_5$, where about half of the electronic
states at the Fermi level have H orbital character.

To understand why in CaBH$_5$ electrons at the Fermi level 
 couple so little to phonons, despite a strong H orbital character, it is useful to inspect
the so-called Local Density of States (LDOS):
\begin{equation}
N(E, \mathbf{r})  = \sum_{n} \int\frac{d^3 k}{(2 \pi)^3}  \delta(E-\varepsilon_{n \mathbf{k}}) 
\left|  \psi_{n \mathbf{k}} (\mathbf{r}) \right|^2,
\end{equation}
where $\psi_n (\mathbf{k})(\mathbf{r})$ and $\varepsilon_{n \mathbf{k}}$ are the Kohn-Sham eigenfunctions and eigenvalues of the system. 
The LDOS 
permits to visualize the real-space distribution of the electronic states at a given energy $E$.
If computed for $E=E_F$,  it permits to visualize the real-space distribution of electrons, which participate in the superconducting pairing.~\cite{Heil_PRB_2018_FeH}

\begin{figure}[h!btp]
\includegraphics[width=0.4\textwidth]{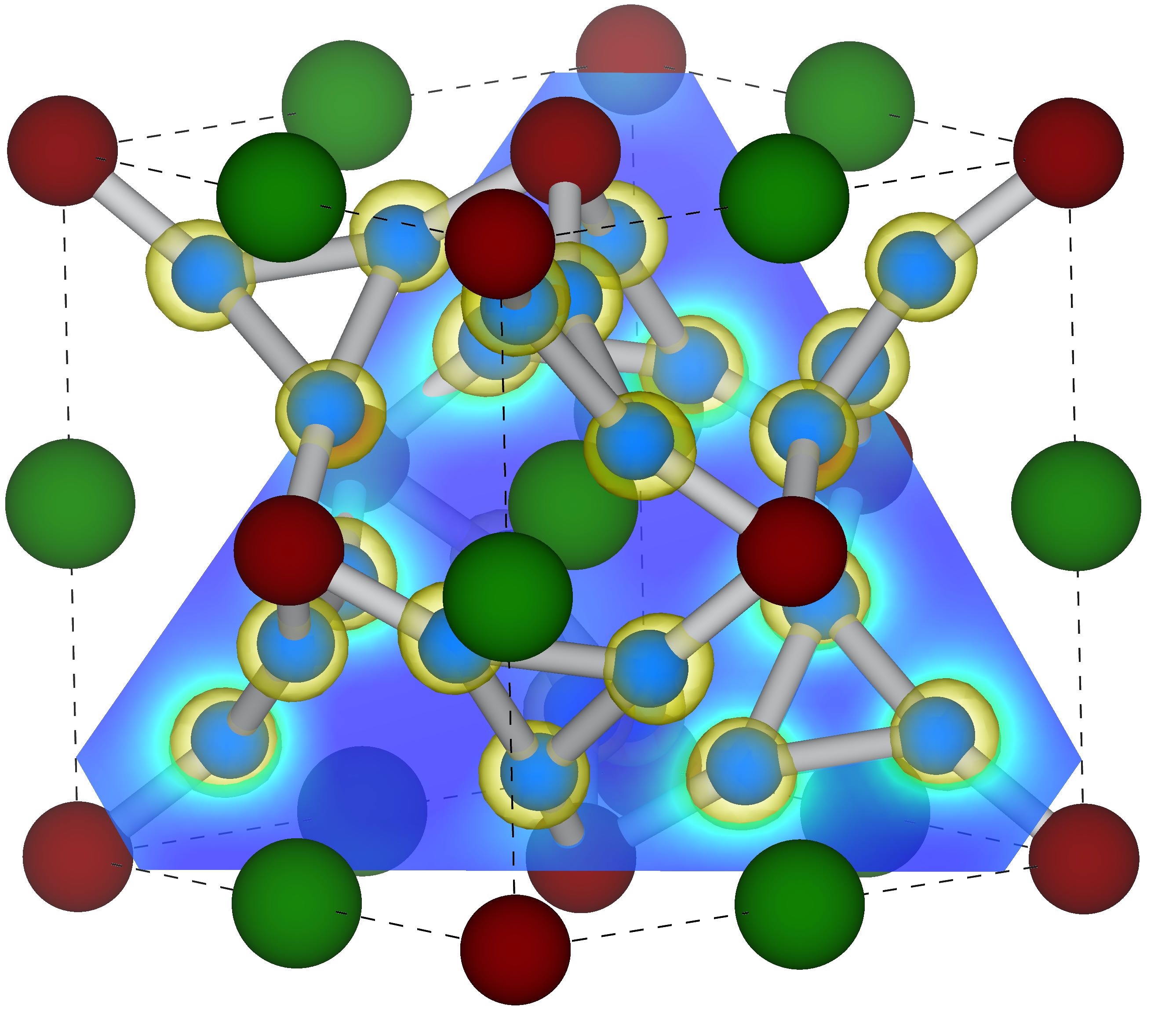}
\includegraphics[width=0.4\textwidth]{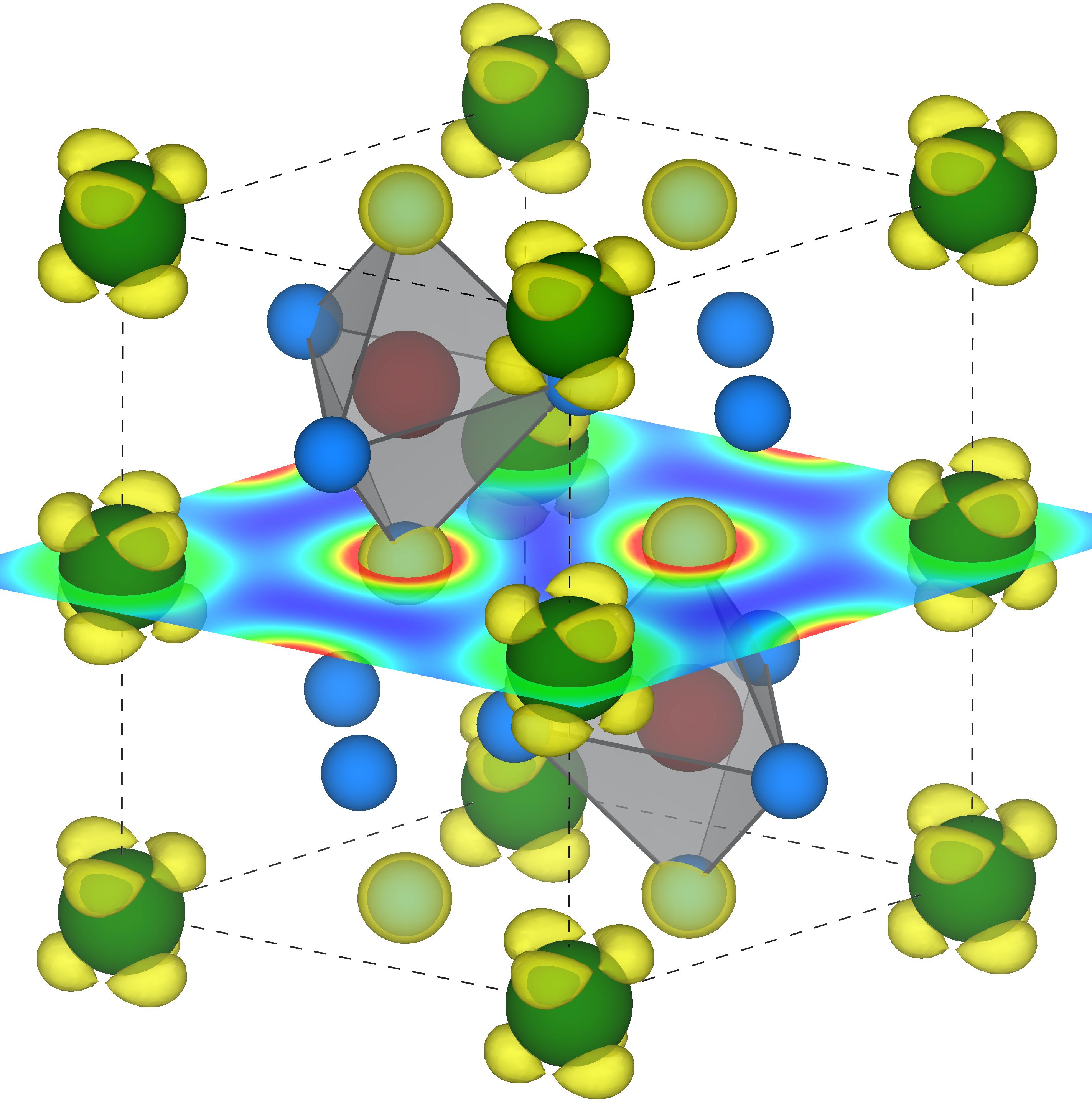}
\caption{LDOS for CaBH$_6$ (top) and   CaBH$_5$ (bottom)The yellow isosurface represents the LDOS distribution at 50\% of its maximum value. The lattice plane shows the LDOS along the 001 plane of CaBH$_5$ and 1$\bar{1}$1 plane of CaBH$_6$. The color scale on the lattice plane goes from the minimum (blue) to the maximum (red) value of the LDOS. The BH$_6$ antiprisms of CaBH$_6$ are not shown for visual clarity.}
\label{fig:LDOS}
\end{figure}

Fig.~\ref{fig:LDOS} shows isosurfaces and isocontours of the LDOS at $E_F$ for  high-\tc{} CaBH$_6$
(top) and non-superconducting CaBH$_5$ (bottom). In CaBH$_6$, the electronic charge
is concentrated on the B-H sublattice, and hence is highly susceptible to B-H bond-stretching vibrations
that dominate the intermediate and high-frequency part of the phonon spectrum, giving rise to large \ep{} matrix elements. In CaBH$_5$, on the other hand, a considerable
fraction of the charge is concentrated around Ca atoms, with a distribution typical of $d$ planar orbitals.
The spatial overlap of these electronic states with
neighboring B or H atoms is quite small, and hence their coupling to B-H vibrations is strongly
suppressed with respect to CaBH$_6$.
The same picture is confirmed by the "fat-band" plots, shown in Appendix \ref{app:supp_figures}, which show that
Ca-3d-derived states dominate one of the two pockets of the Fermi surface of CaBH$_5$, but give none or negligible contributions to the electronic states that form the Fermi surfaces of CaBH$_6$
and Ca$_2$B$_2$H$_{13}$. Hence, Ca $d$ orbitals play an essential role in determining the superconducting properties of different calcium borohydrides.

\section{Conclusions}
\label{sect:conclusions}

In this work, we have studied ab-initio the high-pressure superconducting phase diagram of the ternary calcium-boron-hydrogen system, using evolutionary crystal structure prediction and Density Functional Perturbation Theory, with the aim of identifying new high-\tc{} high-pressure hydrides.
The original hypothesis of this work was that the combination of  hydrogen with a weak (Ca) and a strong (B) hydride former at high pressures could provide a large playground of covalent
metallic crystals, to improve the superconducting properties, compared to
binary hydrides.

Our structural searches have shown that the Ca-B-H phase diagram can be clearly split in two regions: the low-pressure region, extending from room pressure to approximately 100 GPa, is characterized by closed-shell, insulating phases, forming open molecular structures. As pressure is increased, hydrogen is gradually incorporated into the B-H lattice, stabilizing hydrogen-rich compositions. 
At around 100 GPa, a different high-pressure regime sets in, following the rearrangement of boron and hydrogen bonds into dense sublattices; the new structures that form gradually become metallic at higher pressures.
 Among these high-pressure structures, we identify two phases, with unusually high H:B ratios: CaBH$_{6}$ and Ca$_2$B$_2$H$_{13}$, characterized by 5 and 6-coordinated boron atoms, surrounded by hydrogen. For these two phases at 300 GPa, we estimate a \tc{} of 119 and 89 K, respectively, which survive to lower pressures. These \tc{}s are higher than those reported for the stable phases of the binary B-H system \cite{Oganov_PRL_2013_BH}, and comparable to those predicted in the related Li-B-H system.~\cite{Kokail_PRB_2017} 
We also show that  a third hydrogen-rich phase, CaBH$_{5}$, despite being metallic,
is non-superconducting; this can be attributed to a dominant contribution of Ca $d$ states to the electronic states at the Fermi surface, which couple poorly to B and H phonons. In this work, we neglected anharmonic effects which are known to affect superconducting properties of hydrides\cite{Errea_PRL_SH3_2015, Errea_Nature_2020_LaH10} since this level of detail was beyond the scope of our exploratory work.

On one hand, our results confirm that ternary hydrides are a promising venue to search for high-temperature conventional superconductivity; on the other hand, they also highlight that details of electronic structure and chemistry may play a key role in determining the final superconducting \tc.
In fact, we find that high-pressure ternary Ca-B-H covalent metallic structures are not competitive with  the best covalent hydrides, such as \sh, for two reasons: 1) since boron is very electronegative, closed-shell, insulating structures survive up to high pressures, and metalization slowly sets in only after 100 GPa; 2) Since the $d$ orbitals of Ca lie close to the valence, at high pressure they may influence the behavior of valence electronic states in unpredictable ways, acting very differently from other alkali metals and alkaline earths.
Our study hence suggests that searches for high-\tc{} superconductivity in ternary hydrides should be oriented toward compounds containing elements less electronegative than boron, such as aluminum and silicon, where bonds can be more easily broken and reformed.

The authors acknowledge computational resources from the dCluster of the Graz University of Technology and the VSC3 of the Vienna University of Technology, and support through the FWF, Austrian Science Fund, Project P 30269- N36 (Superhydra).  L. B. acknowledges funding through Progetto Ateneo Sapienza 2017-18-19 and computational Resources from CINECA, proj. Hi-TSEPH.

\appendix

\section{Supplementary Figures}
\label{app:supp_figures}

In this section we report the density of states for the stable phases of the Ca-B-H system at 0, 50, and 100 GPa,
discussed in Sect.~\ref{sect:elec_struct}, as well as the orbital-projected band structure plots for
CaBH$_5$, CaBH$_6$, Ca$_2$B$_2$H$_{13}$ discussed in Sect.~\ref{sect:supercon}.

\subsection*{Electronic Densities of States}

\begin{figure}[htpb] 
	\centering
	\includegraphics[width = 0.45\textwidth]{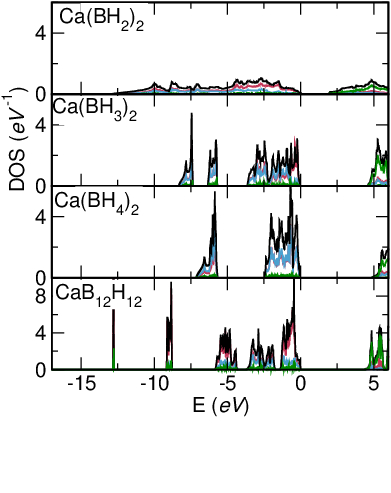}
	\caption{Projected DOS for the following calcium boron hydrides at 0 GPa: the $C2/c$ phase of Ca(BH$_2$)$_2$, the $C2/c$ phase of Ca(BH$_3$)$_2$, and the $Fddd$ phase of Ca(BH$_4$)$_2$. Black, red, turquoise, and green lines represent the total DOS, and its projection onto B, H, and Ca, respectively. The energy scale is shifted with respect to the valence band maximum, the DOS is in units of states per eV per formula unit}
	\label{fig:0GPa_projbands}
\end{figure}
\begin{figure}
	\centering
	\includegraphics[width = 0.45\textwidth]{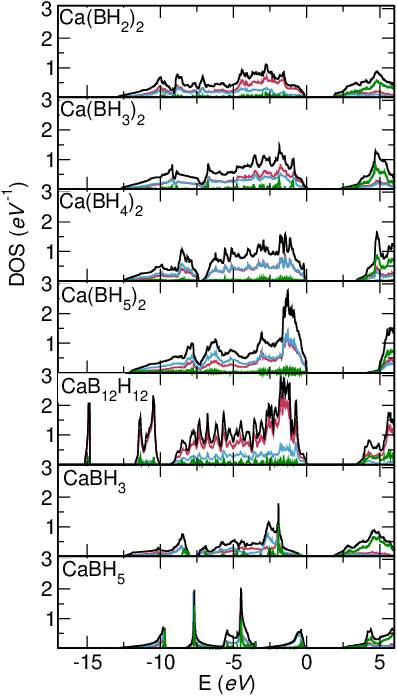}
	\caption{Projected DOS for the following calcium boron hydrides at 50 GPa: the $C2/c$ phase of Ca(BH$_2$)$_2$, the $P2_12_12$ phase of Ca(BH$_3$)$_2$, the $P2_1$ phase of Ca(BH$_4$)$_2$, the $P2_1/m$ phase of CaBH$_3$, and the $F\bar{4}3m$ phase of CaBH$_5$. Black, red, turquoise, and green lines represent the total DOS, and its projection onto B, H, and Ca, respectively. The energy range is shifted with respect to the valence band maximum, the DOS is in units of states per eV per formula unit.}
	\label{fig:50GPa_projbands}
\end{figure}
\begin{figure}
	\centering
	\includegraphics[width = 0.45\textwidth]{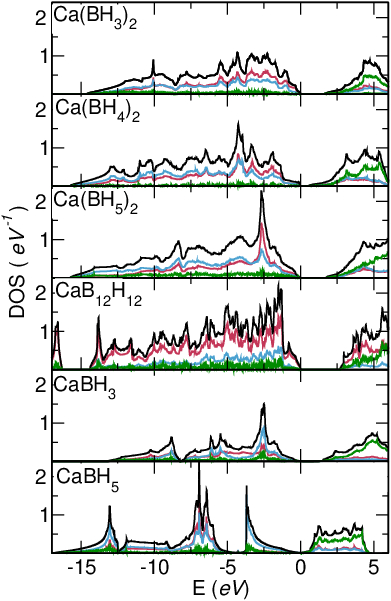}
	\caption{Projected DOS for the following calcium boron hydrides at 100 GPa: the metastable $P2_12_12$ phase of Ca(BH$_3$)$_2$, the metastable $Cmcm$ phase of Ca(BH$_4$)$_2$, the $P\bar{1}$ phase of Ca(BH$_5$)$_2$, the $R\bar{3}m$ phase of CaB$_{12}$H$_{12}$, the $P21/m$ phase of CaBH$_3$, and the $F\bar{4}3m$ phase of CaBH$_5$. Black, red, turquoise, and green lines represent the total DOS, and its projection onto B, H, and Ca, respectively. The energy range is shifted with respect to the valence band maximum (or the Fermi energy, for CaBH$_5$), the DOS is in units of states per eV per formula unit.}
	\label{fig:100GPa_projbands}
\end{figure}

\subsection*{Electronic Band Structure of high-pressure CaBH$_5$, CaBH$_6$ and Ca$_2$B$_2$H$_{13}$}

\begin{figure}[htpb] 
	\includegraphics[width = 0.45\textwidth]{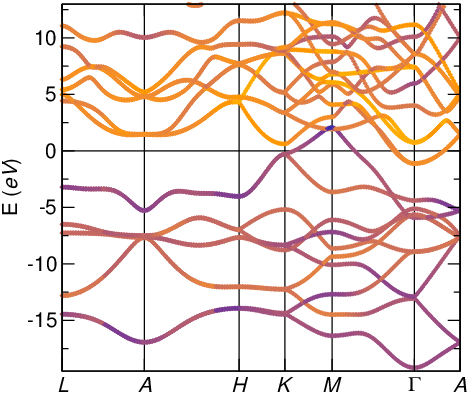}
	\caption{$P6_3mmc$ phase of CaBH$_5$}
	\label{fig:CaBH5_wann_bands}
	\includegraphics[width = 0.45\textwidth]{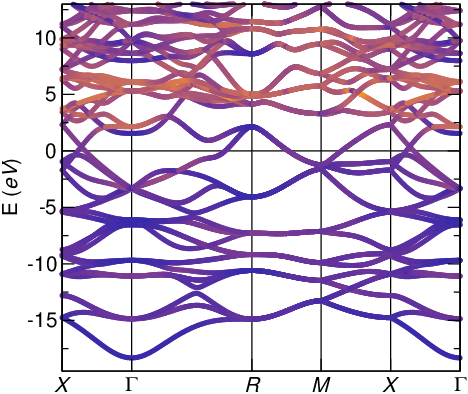}
	\caption{$Pa\bar{3}$ phase of CaBH$_6$}
	\label{fig:CaBH6_wann_bands}
	\includegraphics[width = 0.45\textwidth]{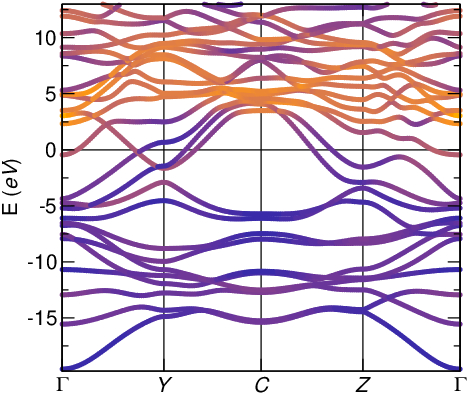}
	\caption{$Pm$ phase of Ca$_2$B$_2$H$_{13}$}
	\label{fig:Ca2B2H13_wann_bands}
	\caption{Band structure projected onto Ca-$d$ states of CaBH$_5$, CaBH$_6$, and Ca$_2$B$_2$H$_{13}$, at 300 GPa. Blue to orange color gradient indicates the fraction of Ca-$d$ character.}
\end{figure}

\section{Computational Details}
\label{app:ComputationalDetails}
\subsection*{Structural Prediction}

To construct the Ca-B-H phase diagram, variable-composition structural searches at fixed pressures were carried out using evolutionary algorithms as implemented in the Universal Structure Predictor: Evolutionary Xtallography software (USPEX) \cite{CompPhysComm_Oganov_2006_USPEX}. Oversampling of the same minima is avoided through the use of a \textit{anti-seeding} technique. Each structure underwent a five-step relaxation to minimize stress and forces, calculated within DFT. For this purpose we employed the Vienna ab-initio software package (VASP) \cite{Kresse_PRB_1996_VASP}, using Projector Augmented Waves (PAW) pseudopotentials with Perdew-Burke-Ernzerhof (PBE) exchange-correlation functional supplied within the VASP package. We used a cutoff on the plane waves expansion of 600 eV; for reciprocal-space integration we employed
a regular grid in k space with a 0.04 spacing in units of $\frac{2\pi}{\text{{\AA}}}$, and a gaussian smearing of 0.05 eV. 

\begin{figure}[htpb]
	\centering
	\includegraphics[width=1\linewidth]{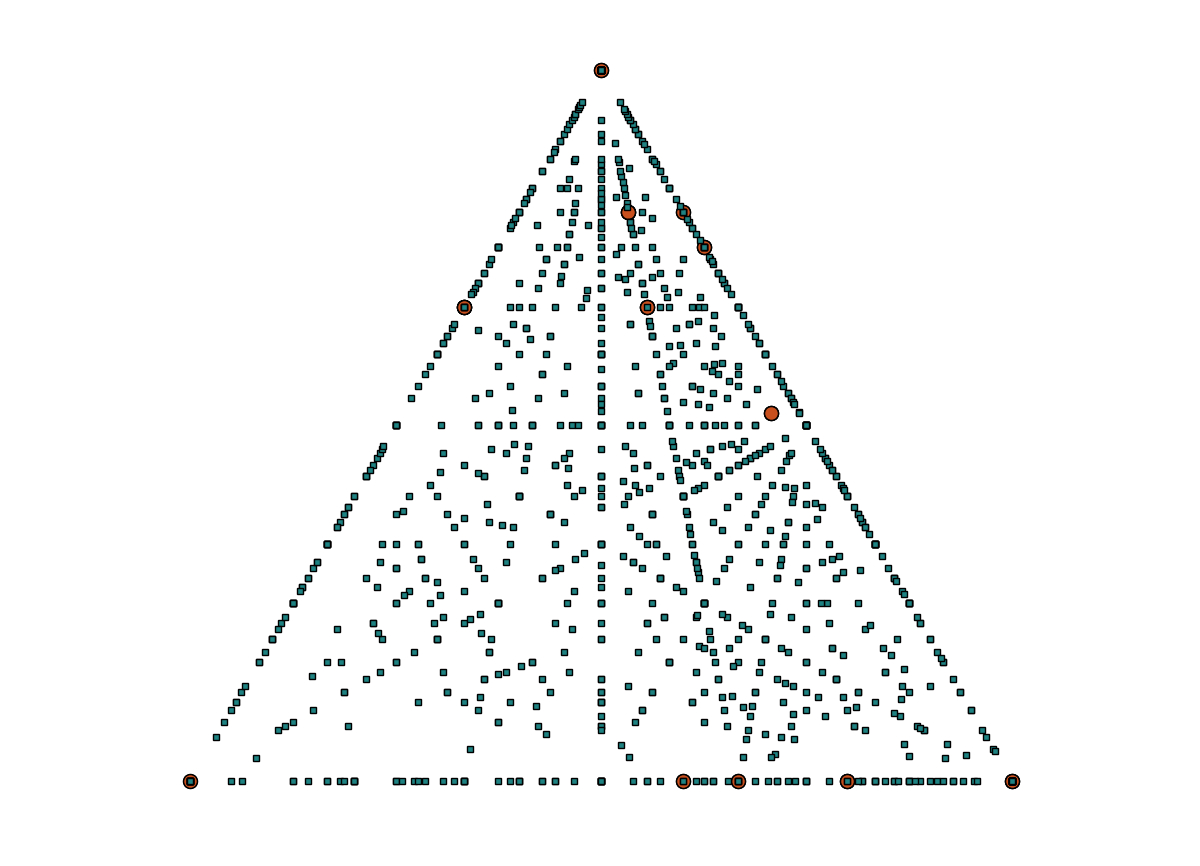}
	\caption{Convex hull (before refinement) for the Ca$_x$B$_y$H$_z$ system at 0 GPa, resulting from exploratory runs over all compositions and over specific lines. A total of about 800 different stoichiometries were sampled with about 4000 structures, amounting to an average of 5 structures/composition.}
	\label{fig:convexhull_all_0GPa}
\end{figure}

The ternary convex hulls in Fig.\ref{fig:convexhulls} were constructed as follows: first, an exploratory run was carried out on the full compositional space, for a total of about 2000 structures per pressure, with a variable number of atoms/cell, ranging from 8 to 16. This was sufficient to correctly predict some stable stoichiometries at room pressure, but not all of them were identified.
To improve the accuracy, we performed additional searches for variable compositions, along selected lines: (Ca$_1$B$_2$)$_{1-x}$H$_x$, (Ca$_2$B$_2$)$_{1-x}$H$_x$, (Ca$_2$B$_1$)$_{1-x}$H$_x$, and Ca$_{1-x}$(BH)$_x$, Ca$_{1-x}$B$_x$, Ca$_{1-x}$H$_x$, and B$_{1-x}$H$_x$,
with about 250 structures per line, and a cell size ranging from 8 to 18. This was sufficient to reproduce results from Refs \onlinecite{Kolmogorov_PRB_2013_CaB, Ma_PNAS_2011_CaH6, Oganov_PRL_2013_BH}, and represents a good compromise between exploration and sampling.
As an example, the convex hull at 0 GPa before refinement including all the compositions sampled in the search is shown in Fig.\ref{fig:convexhull_all_0GPa}. Finally, we performed a few fixed-composition structural searches on the compositions which resulted to be stable, to correctly assess the details of the crystal structure.
 Structures in  literature which had a cell size exceeding these parameters were included by hand.

After these exploratory runs, stable or weakly-metastable structures were further relaxed  with tighter constraints, until the individual components of the forces were less than 1 meV/\AA;
finally, the total energy was computed using the tetrahedron method for k-space integration, with a resolution of 0.03 $\frac{2\pi}{\text{\AA}}$, resulting in a refinement on the estimate of the energies.
From the enthalpies thus calculated, the convex hulls, and the figures, were constructed using the Pymatgen library \cite{Pymatgen}.

The pressure-dependent phase diagram in Fig.\ref{fig:phase_diagrams_bars} was obtained calculating the energy versus volume relation for each stable and metastable phase
identified at 0, 50, 100, and 300 GPa; the obtained relation was then  fitted using the Birch-Murnaghan equation of state.
The phase transitions within a single composition were determined by directly comparing the enthalpy thus estimated.
The decomposition pressure for a given stoichiometry was estimated comparing the enthalpy of the reactant $r$ with that of the products $i$, according to the formula:
\begin{equation}
\Delta H = H[Ca_{x_r}B_{y_r}H_{z_r}] - \sum_{i = 1}^{N} \alpha_{i} H[Ca_{x_i}B_{y_i}H_{z_i}]
\end{equation}
Where $\alpha_i$ is the relative fraction of product $i$ in the reaction; whence if $\Delta H$ is negative, the phase is stable, as the enthalpy of the reactant is less than the enthalpy of the products,
considering as possible decomposition products the phases stable  on the convex hull calculated at the closest pressure. 
Where $\alpha_i$ is the relative fraction of product $i$ in the reaction; whence if $\Delta H$ is negative, the phase is stable, as the enthalpy of the reactant is less than the enthalpy of the products.

\subsection*{Superconductivity}
All electronic structure and superconductivity calculations in sections \ref{sect:elec_struct} and \ref{sect:supercon} were carried out with QUANTUM Espresso (QE) using Norm-conserving (NC) pseudopotentials, and Perdew-Burke-Ernzerhof (PBE) exchange-correlation functional. A cutoff of 80 Ry was used for the plane-wave expansion of the wave functions.  The structures were prerelaxed in QE until each component of the forces acting on single atoms was less than 2 meV/\AA. Calculations of the ground-state charge density were carried out using a 0.06 Ry smearing and a $6\times6\times6$ grid in reciprocal space for k space integration. Phonon calculations were performed on a $2\times2\times2$ reciprocal-space grid, and the integration of the electron-phonon matrix element on the Fermi surface was carried out using a $24\times24\times24$ grid, and a gaussian smearing with a width of 270 meV to describe the zero-width limit of the electronic $\delta$ functions.

The phonon density of states was obtained by performing Fourier interpolation on a $20\times20\times20$ grid.

The figures of the crystal structures were generated using VESTA (\textbf{V}isualization for \textbf{E}lectronic and \textbf{ST}ructural \textbf{A}nalysis) \cite{Izumi_JAPPC_2011_VESTA}

\bibliographystyle{apsrev4-1}

\begin{thebibliography}{76}
\bibitem [{pta()}]{ptable_footnote}%
	\BibitemOpen
	The periodic table has 118 elements, therefore there are $\displaystyle {118 \choose 2} = 6903$ possible combinations
	\BibitemShut {NoStop}%
\end{thebibliography}

\end{document}